\documentclass[runningheads]{cl2emult}

\usepackage{makeidx}  
\usepackage{graphicx} 
\usepackage{subeqnar} 
\usepackage{multicol} 
\usepackage{cropmark} 
\usepackage{lnp}      
\makeindex            

\begin{document}
\title*{Extragalactic Spectroscopy with SIRTF/IRS}
\toctitle{Extragalactic Spectroscopy with SIRTF/IRS}
%
%
\titlerunning{Extragalactic Spectroscopy with SIRTF/IRS}
%
\author{Bernhard Brandl\inst{1}
\and Vassilis Charmandaris\inst{1}
\and Keven Uchida\inst{1}
\and Jim Houck\inst{1}}
\authorrunning{Brandl et al.}
%
%
\institute{Cornell University, Ithaca NY 14853, USA}

\maketitle              

\begin{abstract}
The Infrared Spectrograph (IRS)\index{IRS} is one of the three
instruments on board the Space Infrared Telescope Facility
(SIRTF)\index{SIRTF} to be launched in December 2001. The IRS will
provide high resolution spectra (R$\approx$600) from 10--37\,$\mu$m
and low resolution spectra (R$\ge$60) from 5--40\,$\mu$m.  Its high
sensitivity and ``spectral mapping''-mode make it a powerful
instrument for observing both faint point-like and extended sources. 
We discuss
the performance of the IRS on faint extragalactic targets and present
simulated spectra of starbursts and AGNs at high redshift. In addition,
we discuss the determination of redshifts from the low resolution
spectra.
\end{abstract}

\section{The Infrared Spectrograph}

\begin{figure}[!ht]
\centering
\includegraphics[width=\textwidth]{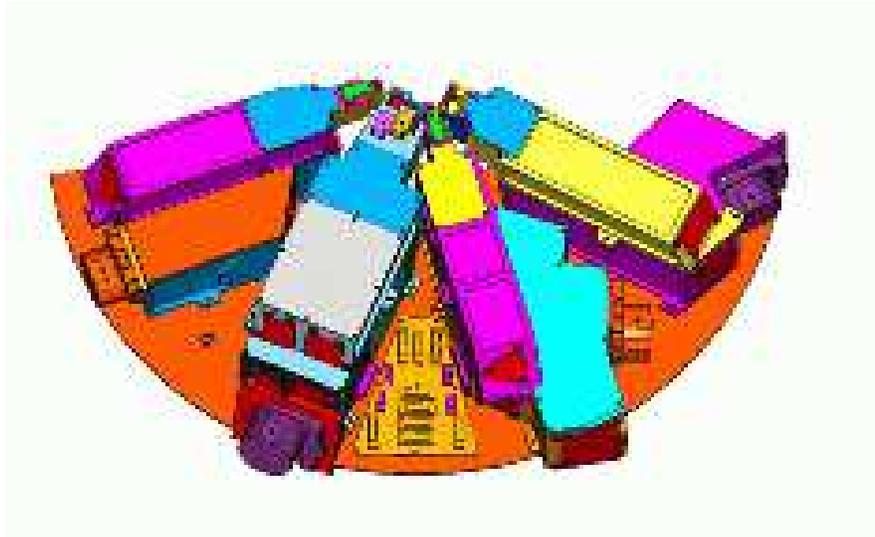}
\caption[]{The 4 IRS modules on their common base plate.}
\label{irs}
\end{figure}

The Infrared Spectrograph (IRS) (Houck \& van Cleve, 1995) will provide 
the Space Infrared Telescope Facility (SIRTF) (Fanson et al. 1998) 
with low and moderate-spectral resolution
spectroscopic capabilities from 4 to 40 microns.  The IRS (see
Fig.~\ref{irs}) is composed of four separate modules, with two of the
modules providing R$\sim$50 spectral resolution over 4 to 40 microns
and two modules providing R$\sim$600 spectral resolution over 10 to 37
microns.

\begin{table}[!ht]
\centering
\caption{Basic IRS characteristics}
\renewcommand{\arraystretch}{1.4}
\setlength\tabcolsep{5pt}
\begin{tabular}{cccccc}
\hline\noalign{\smallskip}
Module & Detector & Pixel Size & Slit Size & $\lambda$ & Resolving \\
& (128$\times$128) & (arcsec) & (arcsec) & ($\mu$m) &  Power (R)\\
\noalign{\smallskip}
\hline
\noalign{\smallskip}
Short Low  & Si:As & 1.8 & 3.6$\times$54.5 & 5.3 -- 7.5 & 62-124\\
 ''	   &   ''    &  ''   &    ''             & 7.5 -- 14  &     ''  \\
Long Low   & Si:Sb & 4.8 & 9.7$\times$145.4 & 14 -- 21  & 62-124\\
''	   &   ''    &   ''  &      ''            & 21 -- 40  &     ''  \\
Short High & Si:As & 2.4 & 4.8$\times$12.1  & 10 -- 19.5 & 600\\
Long High  & Si:Sb & 4.8 & 9.7$\times$24.2  & 19 -- 37  & 600\\
\hline
\end{tabular}
\label{Table1}
\end{table}

The IRS instrument has no moving parts (``bolt-and-go'' philosophy). 
Each module has its own entrance slit in the focal plane.  
The low-resolution modules employ long
slit designs that allow both spectral and one-dimensional spatial
information to be acquired simultaneously on the same detector
array. Two small imaging sub-arrays (``peak-up cameras'') in the 
so-called short-low module
(SL) will also allow infrared objects with poorly known positions to
be accurately placed into any of the IRS modules' entrance slits. The
high-resolution modules use a cross-dispersed echelle design that
gives both spectral and spatial measurements on the same detector array.

\section{The IRS Sensitivity}

\begin{figure}[!ht]
\centering
\includegraphics[width=\textwidth]{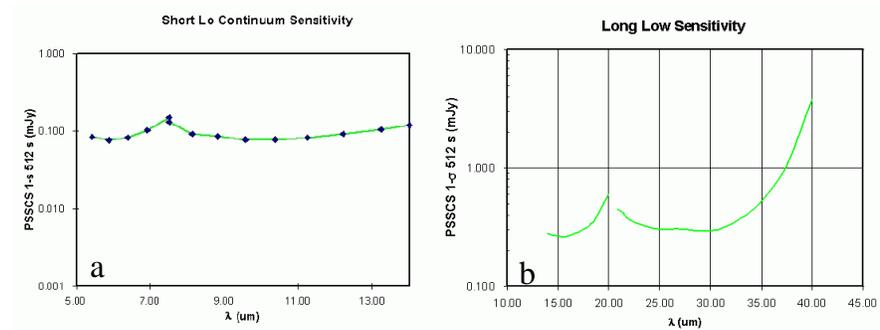}
\caption[]{The theoretical sensitivity plots of the IRS short module
(as of 10/99).}
\label{short}
\end{figure}

\begin{figure}[!ht]
\centering
\includegraphics[width=\textwidth]{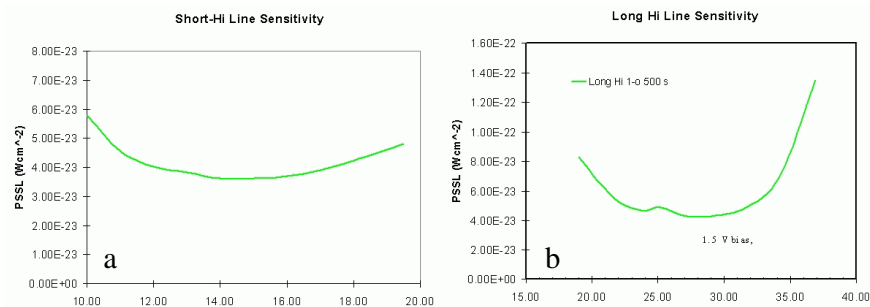}
\caption[]{The theoretical sensitivity plots of the IRS long module 
(as of 10/99).}
\label{long}
\end{figure}

The expected sensitivity of IRS is nearly 10 to 100 times better than that
of the Infrared Space Observatory. The theoretical sensitivity plots for 
the four modules are presented in figures~\ref{short} and \ref{long}.

The continuum point source (5$\sigma$, 500 seconds)
sensitivity of the low resolution module is 1~mJy at 10\,$\mu $m.
The line sensitivity of the high resolution module is 
3$\times$10$^{-18}$~Wm$^{-2}$ at 15\,$\mu $m. The two peak-up cameras
have a sensitivity of about 0.5~mJy.

The saturation limits in 8 seconds for point (extended) sources are 5~Jy
(0.4~Jy\,arcsec$^{-2}$) at 10\,$\mu $m for the low resolution module and
50~Jy (2.1~Jy arcsec$^{-2}$) at 15\,$\mu $m for the high resolution
module. The 4~s saturation limits for the peak-up cameras are 0.5~Jy for 
point sources and 40~mJy\,arcsec$^{-2}$ for extended sources.

\section{Extragalactic Science}

Two of the main areas of extragalactic research where the IRS will be able to
make substantial contributions will be the mid-IR deep surveys as well
as the study of the properties of Luminous Infrared Galaxies (LIRGs).
LIRGs have been studied since the early 70s, but
their importance became evident in 1983 when IRAS revealed tens of
thousands of such infrared galaxies galaxies (i.e., Sanders \& Mirabel
1996). More recently ISO mid-IR deep surveys (i.e., Elbaz et al. 1999)
have shown that the number of galaxies in the distant Universe exceeds
model estimates which were derived from optical observations. These
mid-IR surveys suggest a strong evolution in galaxy formation, such
that more - and perhaps more luminous - galaxies were formed at
earlier epochs.

\begin{figure}[!ht]
\centering
\includegraphics[width=\textwidth]{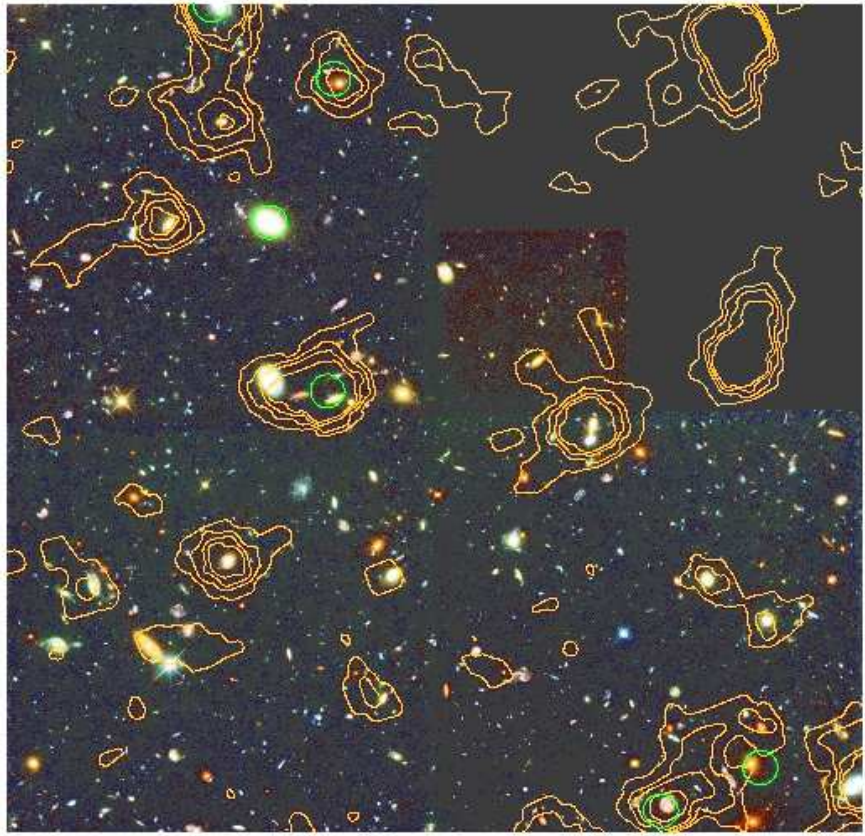}
\caption[]{The HST Hubble Deep Field North with an overlay of the 
ISOCAM 15\,$\mu$m contours (yellow lines). The location of 7\,$\mu$m
ISOCAM detections is indicated with green circles (Aussel et al. 1999).}
\label{hdf}
\end{figure}

As one can clearly see in Fig.~\ref{hdf}
the optical identification of the galaxies with detected mid-IR
emission is ambiguous since the spatial resolution in the
mid-IR is about an order of magnitude inferior to what can be 
achieved in the optical. 

\begin{figure}
\centering
\includegraphics[width=\textwidth]{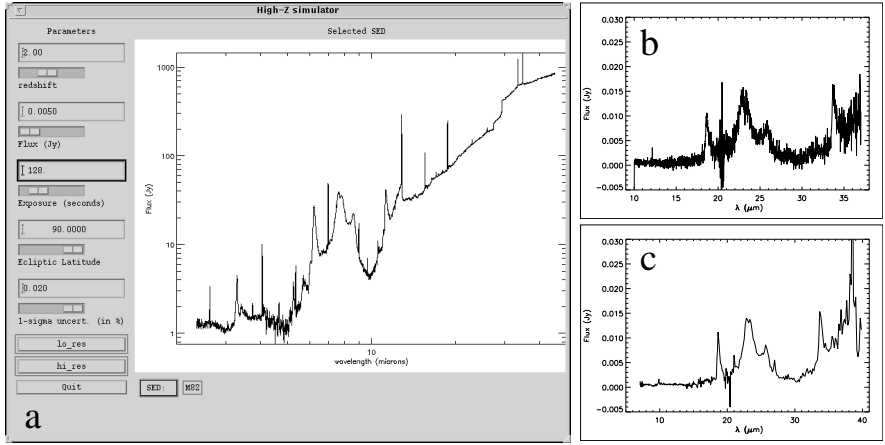}
\caption[]{a) The IRS high-z simulator using as a template the ISO/SWS 
spectrum of M82. b) The expected IRS high-res spectrum of a starburst with an
IRAS 25$\mu$m flux of 5mJy located at $z=2$ after an integration time of
128 seconds. c) The simulated IRS low-res spectrum of the same source.}
\label{m82}
\end{figure}

\begin{figure}
\centering
\includegraphics[width=\textwidth]{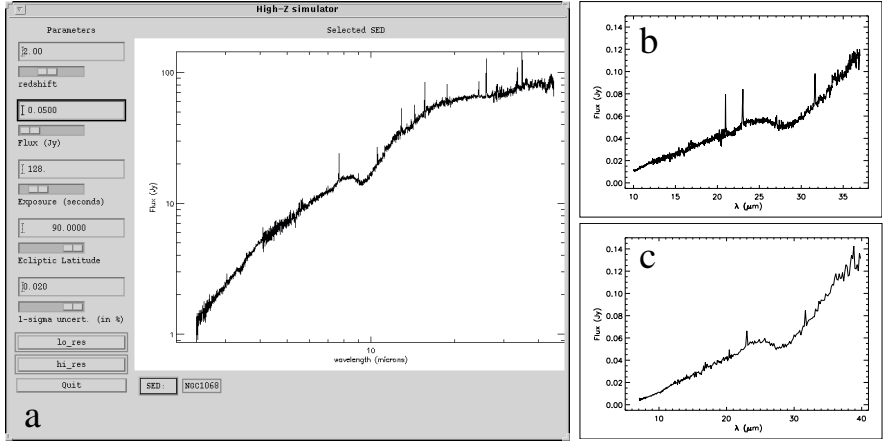}
\caption[]{a) The IRS high-z simulator using as a template the ISO/SWS 
spectrum of NGC1068. b) The expected IRS high-res spectrum of an AGN with an
IRAS 25$\mu$m flux of 5mJy located at $z=2$ after an integration time of
128 seconds. c) The simulated IRS low-res spectrum of the same source.}
\label{n1068}
\end{figure}

The combination of superior sensitivity and good spatial resolution of the
IRS will make it possible to obtain $5 - 40 \mu$m low-resolution mid-IR 
spectra of faint sources in relatively short integration times.  The 
distinct spectrum of the mid-IR emission features at 6.2, 7.7 8.6 and 
11.3~$\mu$m, as well as the presence of the silicate absorption band at 
9.7$\mu$m (restframe wavelengths), can be used to directly derive redshifts
of individual sources as faint as 0.5\,mJy. For illustration we show the 
simulated IRS spectra of a prototypical starburst (M\,82) and active galactic
nucleus (NGC\,1068) at a redshift of 2 in Fig.~\ref{m82} and 
Fig.~\ref{n1068}.
Please note that the fluxes of the template spectra have been normalized to 
user-specified values which may differ from the fluxes of M\,82 and 
NGC\,1068 if they were at a redshift of $z=2$.  Also
note that the specified time is the integration time per module while the 
full $5 - 40 \mu$m is covered by two modules with two settings each.

\section{Conclusions}

The approach presented here provides an unbiased method of deriving directly
the number counts of obscured star forming galaxies as a function of redshift
to a few tenths of a mJy.  Moreover, deep imaging surveys 
with SIRTF or even beyond will reveal the mid/far-IR colors that can be used 
for photometric redshift determinations of the faintest IR galaxies.  The
IRS observations will provide an important calibration sample for these 
surveys. 

More information on the IRS can be found at the following Web-sites:\\
http://astrosun.tn.cornell.edu/SIRTF/irshome.htm\\
http://sirtf.caltech.edu/Observing/obs\_frame.html.

\vspace{5mm}

\noindent{\bf \em Acknowledgements}\\
We'd like to thank Dan Weedman and Jeffrey Wolovitz for providing the basis
of the IRS simulator program, Jeffrey van Cleve for the sensitivity plots
and Eckhard Sturm for providing the ISO-SWS data on M\,82 and NGC\,1068 
prior to publication.

\clearpage
\addcontentsline{toc}{section}{Index}
\flushbottom
\printindex

\end{document}